\documentclass[aps,pra,twocolumn,amsmath,amssymb,showpacs,superscriptaddress]{revtex4-1}
\usepackage{color}
\usepackage{textcomp} 
\usepackage{mathrsfs,amsmath}
\usepackage{graphicx}
\usepackage{dcolumn}
\usepackage[mathlines]{lineno}
\usepackage{hyperref}
\usepackage{bm}
\usepackage{epstopdf}
\usepackage{soul}
\usepackage{epsfig}
\usepackage{bbold}

\usepackage{booktabs}

\usepackage{changes}
\usepackage{lipsum}
\setremarkmarkup{(#2)}

\usepackage{color}

\begin{document}

\title{Observation of local entanglement oscillation in free space}

\author{Eileen Otte}
\affiliation{Institute of Applied Physics, University of Muenster, Corrensstr. 2/4, D-48149 Muenster, Germany}
\author{Carmelo Rosales-Guzm\'an}
\email[Corresponding author: ]{carmelo.rosalesguzman@wits.ac.za}
\affiliation{School of Physics, University of the Witwatersrand, Private Bag 3, Wits 2050, South Africa}
\author{Bienvenu Ndagano}
\affiliation{School of Physics, University of the Witwatersrand, Private Bag 3, Wits 2050, South Africa}
\author{Cornelia Denz}
\affiliation{Institute of Applied Physics, University of Muenster, Corrensstr. 2/4, D-48149 Muenster, Germany}
\author{Andrew~Forbes}
\affiliation{School of Physics, University of the Witwatersrand, Private Bag 3, Wits 2050, South Africa}
\date{\today}

\begin{abstract}
\noindent
It is well known that the entanglement of a quantum state is invariant under local unitary transformations. It dictates, for example, that the degree of entanglement of a photon pair in a Bell state remains maximally entangled during propagation in free-space. Here we outline a scenario where this paradigm does not hold.  Using local Bell states engineered from classical vector vortex beams with non-separable degrees of freedom, so-called classically entangled states, we demonstrate that the entanglement evolves during propagation, oscillating between maximally entangled (purely vector) and product states (purely scalar). We outline the theory behind these novel propagation dynamics and confirm the results experimentally. Crucially, our approach allows delivering a tunable degree of local entanglement to a distant receiver by simply altering a modal phase delay holographically, or, in essence, a tractor beam for entanglement. This demonstration highlights a hitherto unnoticed property of classical entanglement and offers at the same time a device for on-demand delivery of vector states to targets, e.g., for dynamic laser materials processing as well as switchable resolution within STED systems. 

\end{abstract}

\maketitle
\section*{Introduction}
Under local unitary operations, e.g. when propagating through a unitary channel, the degree of entanglement does not change. This is true for both non-local entanglement, i.e., between two photons that are physically separated, and for local entanglement, i.e., between the internal degrees of freedom of a single photon.  Recently, it has become topical to study the latter, and to mimic the former, using vector states of classical light \cite{Spreeuw1998,Pereira2014,Guzman-Silva2015, Karimi2015,Karimi2010,Galvez2012}.  This is possible because the central feature of entanglement, non-separability, is not limited to quantum systems: classical vector beams are likewise non-separable, e.g. in polarisation and spatial modes.  Though the question if such fields can be named ``classically entangled'' is an ongoing one\,\cite{Spreeuw1998, Karimi2015}, practically this property has been exploited for real-time quantum error correction, communication \cite{Souza2008,vallone2014, Milione2015e,Li2016,Milione2015,Ndagano2017}, laser materials processing\,\cite{Nivas2017,Niziev1999,Meier2007}, and metrology\,\cite{Toppel2014,Berg-Johansen2015,DAmbrosio2013}. Also, in imaging \cite{Biss2006,Zhan2009,Chen2013,Segawa2014b}, where tightly focused radially polarized fields are known to produce the narrowest spot size\,\cite{Dorn2003,Youngworth2000,Zhan2002,Lerman2010}, classcially entangled light fields allow super resolution microscopy techniques\,\cite{Torok2004,Hao2010}.  
	  
Here, we demonstrate that entanglement can evolve during propagation in free-space using classically entangled vector vortex beams, non-separable in orbital and spin angular momentum.  We engineer superpositions of these beams such that the entanglement dynamically changes upon propagation from fully entangled (completely non-separable) to no entanglement (fully separable), and by a phase adjustment we show transport of entanglement, reminiscent of tractor beams for particle transport\,\cite{Novitsky2011,Brzobohaty2013,Ruffner2012,Gorlach2017}.  In this way, the degree of entanglement at a receiver may be altered on demand.  The realisation may open new avenues in quantum and classical communication, as well as in improved materials processing (where vector beams and scalar polarised beams are crucial) and enhanced, switchable imaging in e.g. STED microscopy.

\begin{figure*}[t]
\centering
\includegraphics[width=\textwidth]{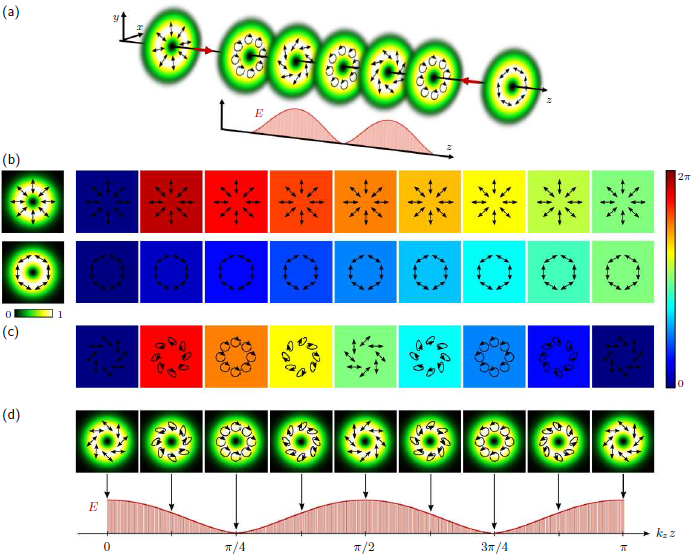} 
\caption{Schematic representation of the generated field with a $z$-dependent degree of entanglement. (a) concept, (b) phase change of radial/ azimuthal beam (top/ bottom) relative to initial phase, (c) absolute value of relative phase difference between radial and azimuthal beam, (d) change in polarisation upon intensity (top) with corresponding degree of entanglement $E$ (bottom) for superimposed counter propagating radial and azimuthal vector beam, all depending on propagation distance $z$ ($k_zz \in [0,\,\pi]$). Further, (b) and (c) include the respective polarisation distribution per distance. \label{fig:fig0conceptnew}} 
\end{figure*}

\section*{Materials and Methods}\label{sec:Results}
{\textbf{Concept.}} Consider a vector beam comprising a superposition of two orthogonally polarised Laguerre-Gaussian modes $|LG_{p}^l(x,y)\rangle$ given by\,\cite{Galvez2012}
\begin{equation}
|\Psi_{\text{VB}}^{\pm}\rangle = \frac{1}{\sqrt{2}}\left(\text{e}^{\text{i}\alpha}\,|LG_{p_1}^{l_1}\rangle\, |R\rangle+ \text{e}^{-\text{i}\alpha}\,|LG_{p_2}^{l_2}\rangle\,|L\rangle\right)\cdot \text{e}^{\pm\text{i}k_zz},
\label{eq:VectorBeams} 
\end{equation} 
\noindent
where we assume a propagation in $\pm z$-direction, approximated by the factor $\text{e}^{\pm\text{i}k_z z}$, with $\vec{k} = (k_x,\,k_y,\,k_z)$ being the wave vector expressed in terms of the wavelength $\lambda$ as $k=~2\pi/\lambda$. The kets $|R\rangle$ and $|L\rangle$ represent the unit vector of right- and left-handed circular polarisation states, respectively, and $\alpha$ defines the phase relation between both states. The indices $l$ and $p$ denote the azimuthal and radial degrees of freedom, respectively, the former being related to the orbital angular momentum (OAM) of the Laguerre-Gaussian beam. In the following description we will restrict ourselves to the case $l_1=-l_2=l$ and $p_1=p_2=p$ but it can be extended to other cases.\\ 
Equation (\ref{eq:VectorBeams}) can be conveniently written as \cite{McLaren2015}
\begin{equation}
|\Psi\rangle = \sqrt{a}\cdot |u_R\rangle|R\rangle + \sqrt{1-a}\cdot |u_L\rangle|L\rangle,
\label{eq:VBQuantumNotation}
\end{equation}
where $|LG_p^l\rangle \cdot\text{e}^{\text{i}(\pm\alpha\pm k_zz)}=|LG_p^l\rangle \,\text{e}^{\text{i}\zeta_{R,L}}$ are represented by the ket $|u_{R,L}\rangle$ and the relative weightings of $|u_R\rangle$ and $|u_L\rangle$ by $a$. Moreover, $|u_{R,L}\rangle$ satisfies the normalisation condition $\langle u_{R,L}|u_{R,L}\rangle$.  

The degree of non-separability (classical entanglement) $E(|\Psi\rangle)\in [0,\,1]$ of a vector field as defined by Eq. (\ref{eq:VBQuantumNotation})  can be computed using tools from quantum mechanics. Here, we consider the entanglement entropy, originally derived for quantum states\,\cite{Wootters2001,Hill1997} and later extended to classical non-separable states\,\cite{McLaren2015} as
\begin{equation}
E(|\Psi\rangle) = -\left[a\cdot \log_2(a)+(1-a)\cdot \log_2(1-a)\right].
\label{eq:VQF1}
\end{equation}
Consequently, if we analyse a vector beam $|\Psi_{\text{VB}}^{\pm}\rangle$ under a unitary transformation, i.e. propagation in free space along $\pm z$-direction (Eq. (\ref{eq:VectorBeams})), where $a=1/2$ for all $z$ values, we observe a spatially invariant degree of entanglement $E(|\Psi_{\text{VB}}^{\pm}\rangle)=1$. \\

Remarkably, we can engineer a light field $|\Psi(x,y,z)\rangle$ with a $z$-dependent degree of entanglement $E(|\Psi\rangle, z)$, by combining two orthogonal vector beams $|\Psi_{\text{VB}_1}^+\rangle$ and $|\Psi_{\text{VB}_2}^-\rangle$, coaxially propagating in opposite directions, as illustrated in Fig. \ref{fig:fig0conceptnew}(a). These fields can be generated by setting $\alpha_{\text{VB}_1}=0$ and $\alpha_{\text{VB}_2}=\pi/2$ in Eq. (\ref{eq:VectorBeams}), namely
\begin{equation}
|\Psi_{\text{VB}_1}^+\rangle=\frac{1}{2}\left(|LG_{p}^{-l}\rangle\,|R\rangle+|LG_{p}^{l}\rangle\,|L\rangle\right)\cdot \text{e}^{\text{i}k_zz}
\label{eq:V1}
\end{equation}
and
\begin{equation}
|\Psi_{\text{VB}_2}^-\rangle=\frac{1}{2}\left(\text{e}^{\text{i}\frac{\pi}{2}}|LG_{p}^{-l}\rangle\,|R\rangle+\text{e}^{-\text{i}\frac{\pi}{2}}|LG_{p}^{l}\rangle\,|L\rangle \right)\cdot \text{e}^{-\text{i}k_zz},
\label{eq:V2}
\end{equation}
\noindent
with a phase distribution as a function of $z$ as shown in Fig.~\ref{fig:fig0conceptnew}(b) top and bottom, respectively, for the case $l=1$, $p=0$. The normalised field resulting from such a superposition takes the form
\begin{align}
|\Psi\rangle = &\frac{1}{2}\left(\text{e}^{\text{i}k_zz}+\text{i}\,\text{e}^{-\text{i}k_zz}\right)\,|LG_{p}^{-l}\rangle|R\rangle \nonumber\\
&+ \frac{1}{2}\left(\text{e}^{\text{i}k_zz}-\text{i}\,\text{e}^{-\text{i}k_zz}\right)\,|LG_{p}^{l}\rangle|L\rangle,
\label{eq:PsiQuantum}
\end{align}
whose polarisation evolution upon propagation for the regarded example is shown in Figs. \ref{fig:fig0conceptnew}(c) and (d), respectively. Hereby (c) includes the change of relative phase between superimposed beams. Now, the degree of entanglement, as defined by Eq. (\ref{eq:VQF1}), for the new vector field $|\Psi\rangle$ is given by
\begin{align}
E(|\Psi\rangle,z) = &1-\frac{1}{2}\left[1+\sin(2k_zz)\right]\cdot \log_2\left[1+\sin(2k_zz)\right]- \nonumber\\
&\frac{1}{2}\left[1-\sin(2k_zz)\right]\cdot \log_2\left[1-\sin(2k_zz)\right]
\label{eq:nonunitaryDegEnt}
\end{align}
(details with respect to calculations can be found within the Supplementary Information). Thus, the state undergoes a periodic variation in the degree of entanglement as function of $z$, as illustrated in Fig. \ref{fig:fig0conceptnew}(d), bottom, while the intensity profile remains constant. Full entanglement, i.e. maximal non-separability, ($E(|\Psi\rangle)=1$) is achieved at $z=n\lambda/4,\,n\in \mathbb{N}$, whereas non-entanglement, i.e. complete separability, ($E(|\Psi\rangle)=0$)  is observed at $z=(2n+1)\lambda/8,\,n\in \mathbb{N}$. Notice that space-variant entanglement of the form $E(|\Psi\rangle,z)$ can be realised by any OAM subspace $l$ by superposition of orthogonal vector fields $|\Psi_{\text{VB}_1}^{+}\rangle$ and $|\Psi_{\text{VB}_2}^{-}\rangle$, as long as they carry the same radial order $p_{1,2}(\text{VB}_{1,2})=p$ (cf. Supplementary Information). \\
This unique property of the field $|\Psi\rangle$ provides a means to facilitate the transport of a chosen degree of entanglement across arbitrary distances, by simply applying a phase adjustment $\varphi$, reminiscent of tractor beams\,\cite{Novitsky2011,Brzobohaty2013,Ruffner2012,Gorlach2017}. To illustrate this, we can replace the propagation factor in Eq. (\ref{eq:VectorBeams}) by the factor $\text{e}^{\pm\text{i}(k_zz+\varphi)}$. In this way, the maximum degree of entanglement ($E_{\text{max}}(|\Psi\rangle)=1$), for example, can be transported to a position $z_{\text{max}}$ according to the expression
\begin{equation}
z_{\text{max}}(\varphi) = \frac{\lambda}{4}\left(m-\frac{2\varphi}{\pi} \right), \,\, m\in \mathbb{Z}.
\label{eq:zmax}
\end{equation} 
This means, any chosen state can be conveyed to a specific position in space, along the propagation axis, by simply adjusting the phase $\varphi$. Moreover, by applying a time dependent phase shift $\varphi(t)$, it is possible to impart a time dependent movement of regarded maximum with an axial velocity given by 
\begin{equation}
v_{\text{max}}(t) = -\frac{\lambda}{2\pi}\frac{\partial \varphi(t)}{\partial t}.
\label{eq:vmax}
\end{equation}

\begin{figure*}[t]
\centering
\includegraphics[width=\textwidth]{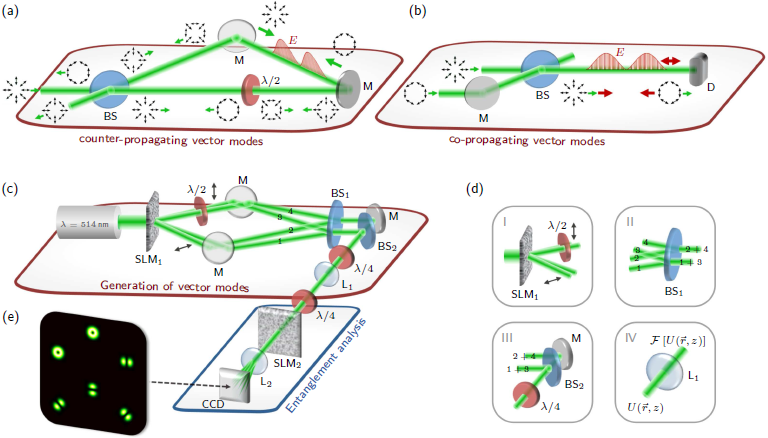} 
\caption{Sketch of experimental concept: Approach of (a) counter- and (b) co-propagating vector modes for the realisation of the light field $|\Psi\rangle$. (c) Applied system for generation (red box) and analysis (blue box, (e)) of $|\Psi\rangle$ with experimental steps indicated in (d). SLM$_{1,2}$: spatial light modulator $|$ $\lambda/2$: half-wave plate $|$ $\lambda/4$: quarter-wave plate $|$ M: mirror $|$ BS$_{1,2}$: beam splitter $|$ L$_{1,2}$: lens $|$ CCD: camera. \label{fig:fig1concept}} 
\end{figure*}

\textbf{Experimental details.} A simple method to generate the desired light field $|\Psi\rangle$, revealing the dynamics of entanglement, is via an interferometric approach. An exemplary system is sketched in Fig. \ref{fig:fig1concept}(a). By combining a Sagnac interferometer with a half-wave plate (diagonally oriented), a single incident vector beam, e.g. radially polarized, can be used for the generation of a standing wave of entanglement $E$ according to Eqs. (\ref{eq:V1}--\ref{eq:PsiQuantum}). Note that in each arm of the interferometer, counter-propagating (green arrows) vector modes of orthogonal polarisation will give rise to a classically entangled standing wave, as indicated by a red curve in one of the arms. \\
Even if this approach of counter-propagating beams is very intuitive, the investigation of the light field $|\Psi\rangle$ would be challenging as any measuring device inserted in the path would destroy the classically entangled field. Here, we propose an alternative approach that allows us to quantify the spatially varying degree of entanglement. This approach is based on the engineered superposition of co-propagating orthogonally polarized vector modes, as visualized in Fig. \ref{fig:fig1concept}(b). By applying digital propagation\,\cite{Goodman1996, Schulze2012a}, we are able to artificially counter-propagate the two modes (red arrows), which physically co-propagate in the same direction (green arrows), enabling us to investigate the realised light field $|\Psi\rangle$ along the beam path.\\ 
The digital propagation of a light field $U(\vec{r}, z)$ propagating in $z$-direction is based on the angular spectrum\,\cite{Goodman1996,Schulze2012a}, according to which $U(\vec{r}, z) = \mathcal{F}^{-1}\left\{\mathcal{F}\left[U(\vec{r}, 0)\right] \cdot \text{e}^{\text{i}k_zz}\right\}$, where, $\vec{r}=(x,y,z)$ is the coordinates in real space and $\mathcal{F}$, $\mathcal{F}^{-1}$ the Fourier and inverse Fourier transforms, respectively. Following this, by the application of Fourier holograms in combination with a phase shift $\pm k_zz$, encoded on a spatial light modulator (SLM), we were able to digitally propagate a light field in $\pm z$-direction. In order to independently control the phase shift of each vector mode, to artificially generate counter-propagating vector modes, we developed a new method that facilitates the generation of any vector beam using a multiplexing approach enable by an SLM. This method allows not only simultaneous generation of multiple vector modes but also their independent manipulation, such as digital propagation.\\ 
The idea behind our method is to encode a superposition of different holograms, each with a different spatial carrier frequency (blazed grating), on a single SLM\,\cite{Rosales2017}. Thus, each beam is sent to different transverse positions in space that allows manipulating their polarisation independently, as required for vector beam generation. For example, to generate a radially polarised vector beam we multiplexed the corresponding holograms to create two helical $LG$-beams with opposite topological charges ($l=\pm 1$) on the SLM. A half-wave plate placed in the path of one beam changes its polarisation from horizontal to vertical. Both beams were then recombined and passed through a quarter-wave plate to change horizontal and vertical polarisations into left- and right-circular polarisations, respectively, generating in this way the desired vector beam. \\
In the present case, where we realised a superposition of two cylindrical vector beams VB$_{1,2}$ (see Fig. \ref{fig:fig1concept}(c), red box, and (d)), four vortex beams were multiplexed in the SLM (SLM$_1$; Fourier holograms), manipulated accordingly and (counter-)propagated digitally (Fourier relation between SLM$_1$ and SLM$_2$ by lens L$_{1}$) to generate the desired field $|\Psi\rangle$ within the observation plane (SLM$_2$). In this way, the detection system may remain static while the created vector beams artificially propagate in opposite directions. Beyond this, applying a time varying phase shift for digital propagation by the SLM facilitates the transport of a chosen degree of entanglement to arbitrary positions similar to the case of tractor beams.\\

\textbf{Theory of entanglement entropy.} For the analysis of the realised light field $|\Psi\rangle$, we determine the degree of classical entanglement, i.e. the degree of non-separability, in different $(x,y)$-planes. An appropriate tool for this is the quantum mechanics entanglement entropy\,\cite{McLaren2015, Wootters2001}
\begin{equation}
E =h\left(\frac{1+s}{2}\right),
\label{eq:EntanglementEntropy}
\end{equation} 
with $h(r) = -r\log_2(r)-(1-r)\log_2(1-r)$. Here, $s$ is the length of the Bloch vector given by $s = \left(\sum_i \langle\sigma_i\rangle^2\right)^{1/2}$ with $i = \{1,2,3\}$. By $\langle\sigma_i\rangle$ the expectation values of the Pauli operators are represented. These values are obtained by a set of 12 normalised, on-axis intensity measurements or six identical measurements for two different basis states\,\cite{Ndagano2016, McLaren2015}. \\
We chose circular polarisation as basis. As a consequence, the projection measurements are given by two modes carrying OAM of topological charge $l$ and $-l$, in addition to four superposition states represented by $\exp(\text{i}l\phi)+ \exp(\text{i}\gamma)\exp(-\text{i}l\phi)$ with $\gamma = \{0,\,\pi/2l,\,\pi/l,\,3\pi/2l\}$ ($\phi$: azimuthal angle in polar coordinates). In the case at hand, we investigate vector modes of first order (cf. results section), hence, projection measurements are performed for $l=1$.\\
 \begin{table}
 \caption{Normalised intensity measurements $I_{uv}$ for the determination of expectation values $\langle \sigma_i \rangle$. \label{tab:Tomography}}
 \begin{ruledtabular}
 \begin{tabular}{c|c c c c c c c}
Basis states & $l=1$ & $l=-1$ & $\gamma =0$ & $\pi/2$ & $\pi$ & $3\pi/2$& \\ 
&\includegraphics[width=0.07\linewidth]{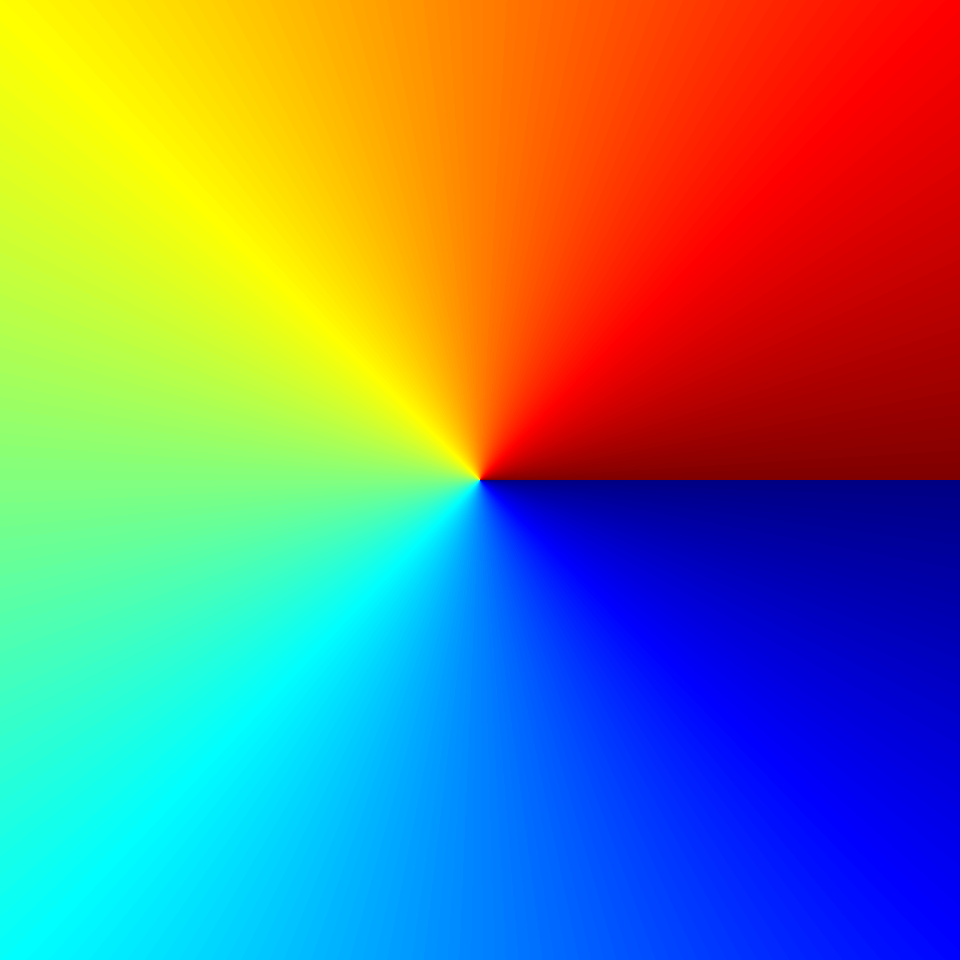}&\includegraphics[width=0.07\linewidth]{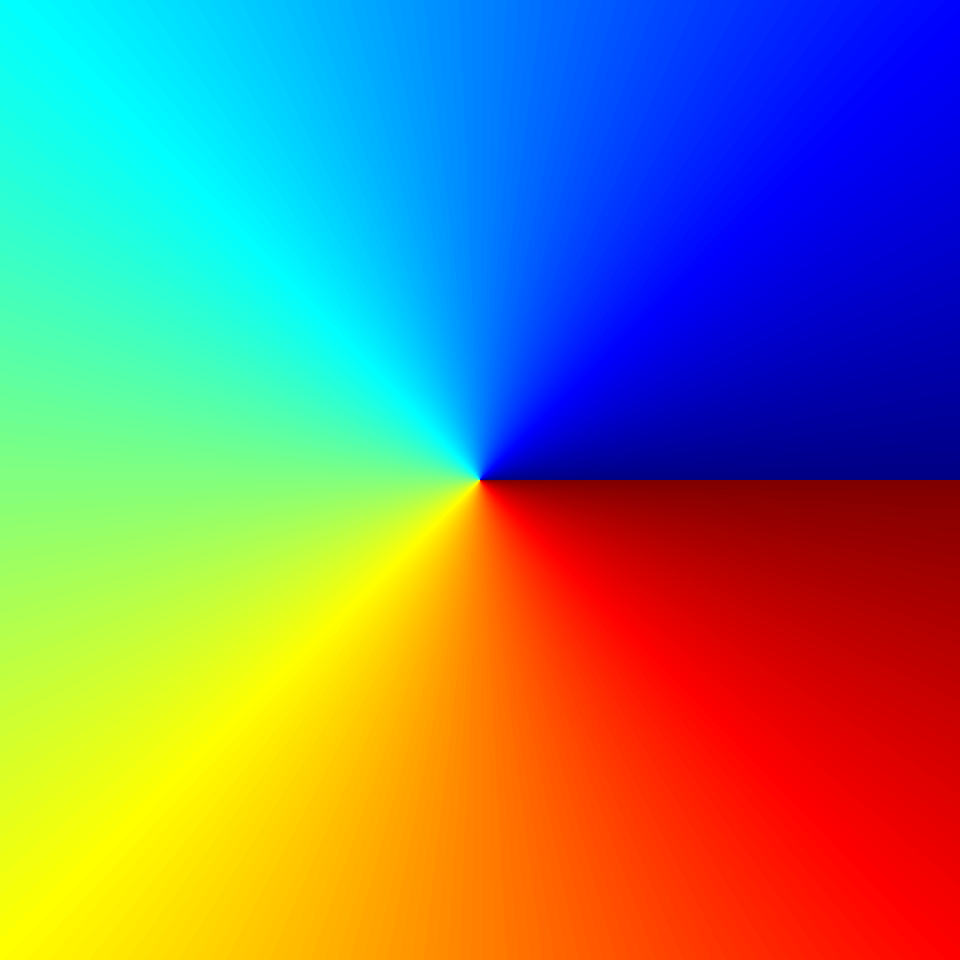}&\includegraphics[width=0.07\linewidth]{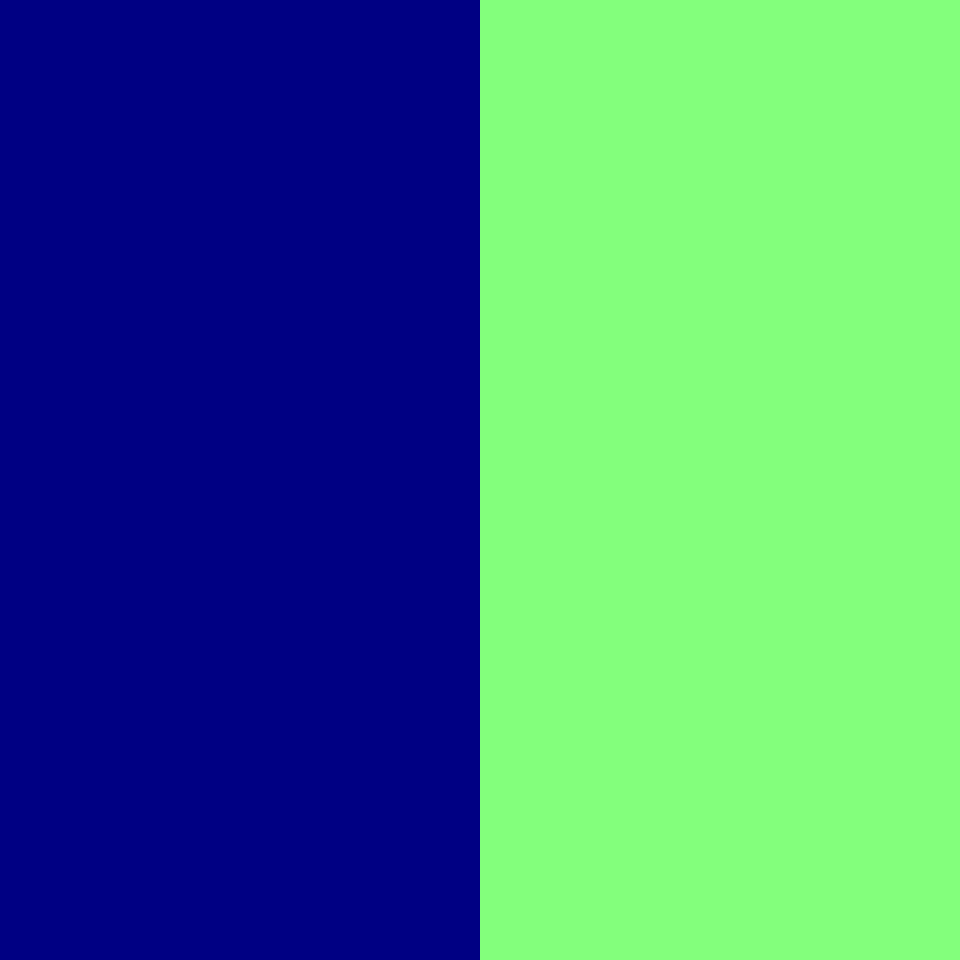}&\includegraphics[width=0.07\linewidth]{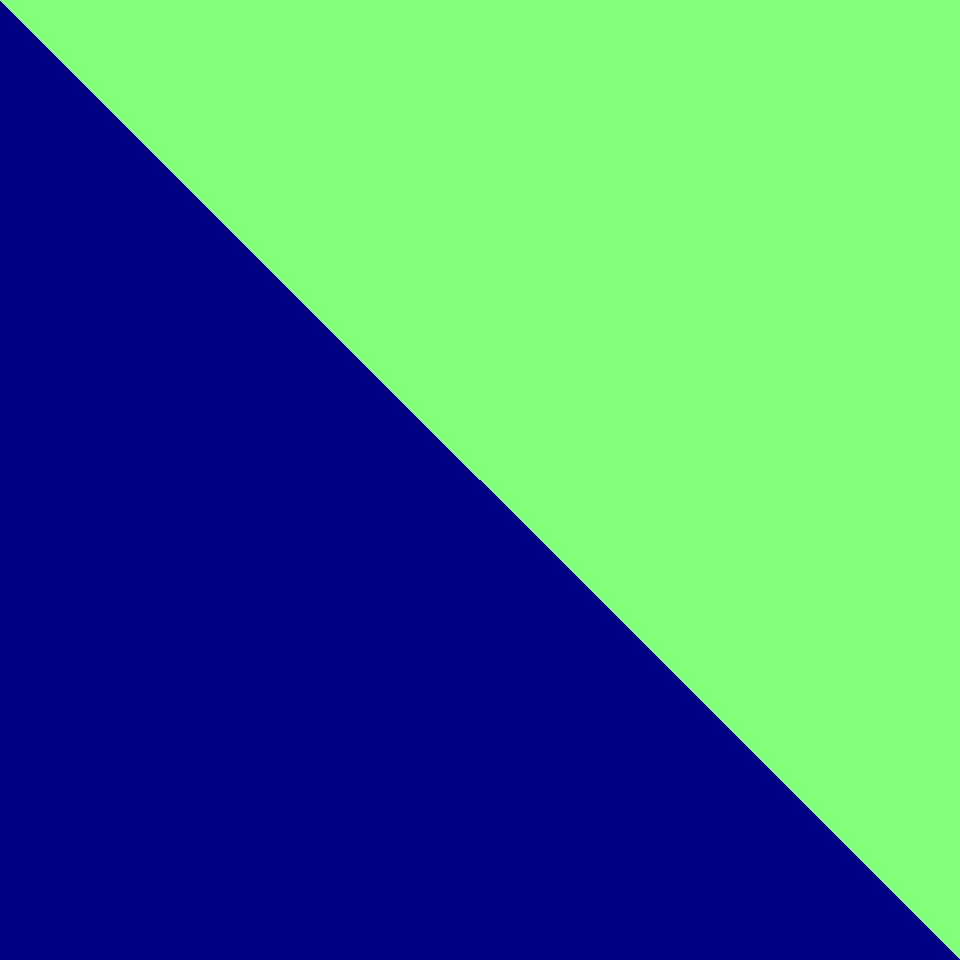}&\includegraphics[width=0.07\linewidth]{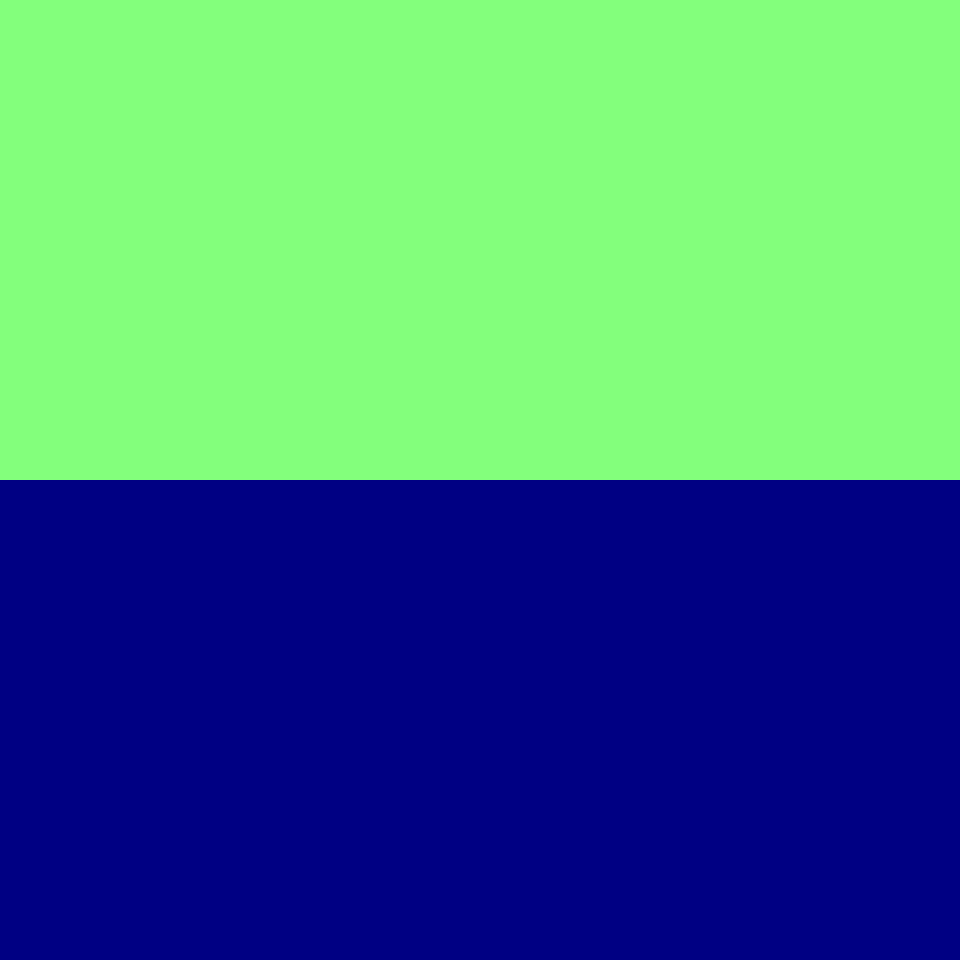}&\includegraphics[width=0.07\linewidth]{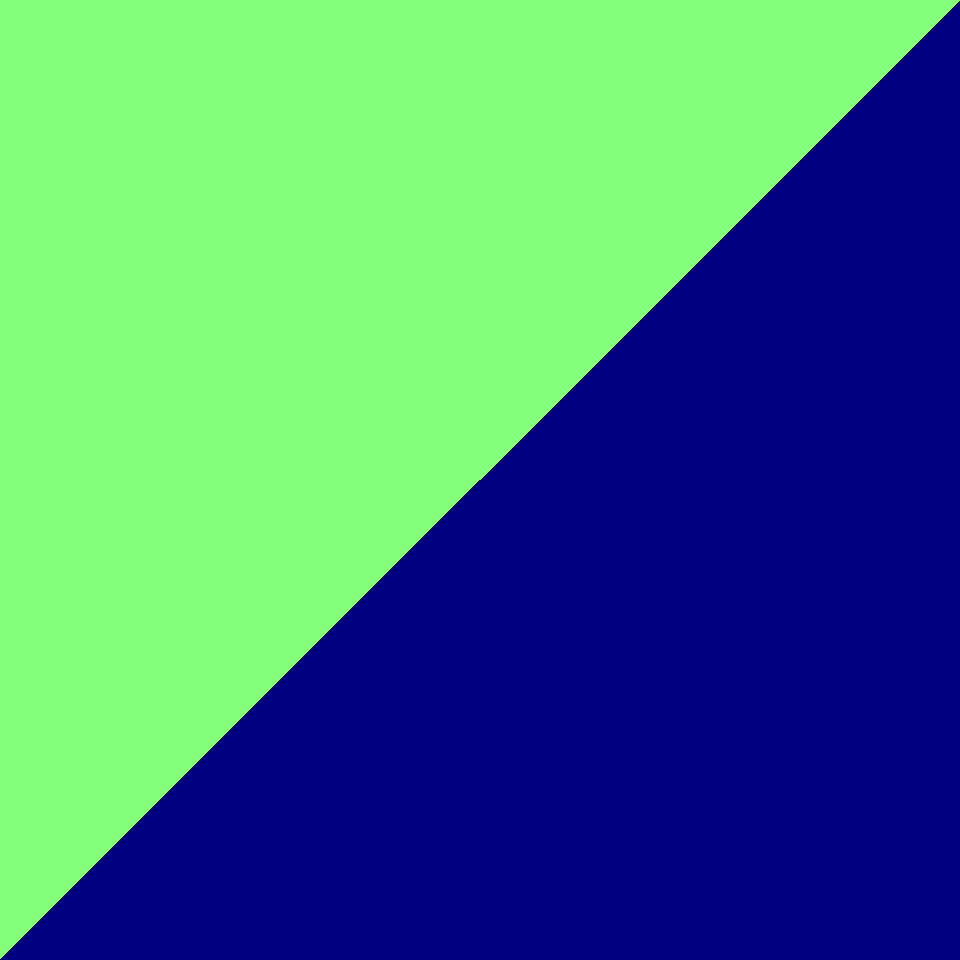}&\includegraphics[width=0.034\linewidth]{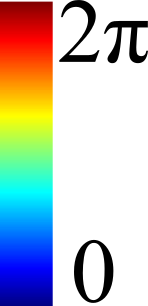}\\ \hline \hline
Left circular $\vert L\rangle$& $I_{11}$ & $I_{12}$ & $I_{13}$ & $I_{14}$ & $I_{15}$ & $I_{16}$ &\\ 
Right circular $\vert R\rangle$& $I_{21}$ & $I_{22}$ & $I_{23}$ & $I_{24}$ & $I_{25}$ & $I_{26}$& \\ 
 \end{tabular}
 \end{ruledtabular}
 \end{table}
According to Table \ref{tab:Tomography}, the expectation values $\langle \sigma_i \rangle$ are calculated from
\begin{align}
&\langle \sigma_1\rangle = (I_{13}+I_{23})-(I_{15}+I_{25}),\\
&\langle \sigma_2\rangle = (I_{14}+I_{24})-(I_{16}+I_{26}),\\
&\langle \sigma_3\rangle = (I_{11}+I_{21})-(I_{12}+I_{22}).
\end{align}
To determine the entanglement entropy $E$ experimentally, we measure the on-axis intensity values $I_{uv}$ with $u,\,v\in \lbrace 1,\,2,\,3\rbrace$, as indicated in Fig. \ref{fig:fig1concept}(c), blue box, and (e). Therefore, polarisation projections are performed by the use of a quarter-wave plate ($\lambda/4$) set to $\pm 45^{\circ}$ in combination with a polarisation sensitive spatial light modulator (SLM$_2$), and OAM projections by a phase pattern on this modulator. The respective phase pattern is carrying the information of all six OAM projections, each of them assigned to another spatial carrier frequency\,\cite{Ndagano2016}. The application of this demultiplexing hologram results in six outputs on the CCD camera (Fig. \ref{fig:fig1concept}(e)) positioned in Fourier relation with the observation plane (SLM$_2$) by a lens (L$_2$), enabling a single-shot measurement for each polarisation basis.\\
For the entanglement entropy analysis in different $(x,y)$-planes of the light field, artificial propagation in $\pm z$-direction is applied. Further, intensities $I_{uv}$ for different planes are normalised by $I_{11}+I_{12}+I_{21}+I_{22}$ for left- and right-circular polarisation basis.
 
\section*{Results and Discussion}
To verify that the field $|\Psi\rangle$, engineered in the described way, follows the entanglement dynamics predicted by Eq. (\ref{eq:nonunitaryDegEnt}), we experimentally generated and superimposed two orthogonal vector beams (according to Eqs. (\ref{eq:V1}) and (\ref{eq:V2})), using the setup shown in Fig. \ref{fig:fig1concept}(c), indicated by the red box. For simplicity but without the loss of generality, we chose first-order radially and azimuthally polarised modes with $l=1$ and $p=0$. Close ups to the different sections of the generation process are shown in (d). The desired light field $|\Psi\rangle$ is realised in the Fourier plane (SLM$_2$, observation plane) of SLM$_1$. \\

\begin{figure}[htbp]
\centering
\includegraphics[width=.5\textwidth]{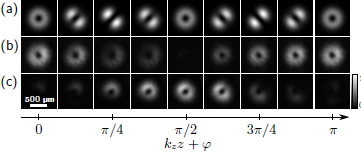} 
\caption{Intensity profile of the engineered light field $|\Psi\rangle$ for various $z$-positions in units of $k_zz+\varphi$ ($\varphi = -\pi/4$). (a) Normalised intensity profile of the field $|\Psi\rangle$, passing through a horizontally aligned polarizer (data from simulation). Experimental results of counter-oscillating intensities for (b) $|L\rangle$ and (c) $|R\rangle$ polarisation components.
\label{fig:fig2prop}} 
\end{figure}

The artificially generated field $|\Psi\rangle$ can be separated into its $|R\rangle$ and $|L\rangle$ parts, each of those including two counter-propagating $LG$-modes of the same helicity. For each polarisation, one mode propagates in positive $+z$-direction and the other one in negative $-z$-direction, achieved through digital propagation enabled by SLM$_1$. The digital propagation was encoded on the SLM as $\exp[\pm\text{i}\,(k_zz+\varphi)]$, whereby we chose $\varphi$ to be a discrete phase offset of $-\pi/4$. Using a CCD camera positioned in the observation plane, we recorded the intensity profile of $|R\rangle$ and $|L\rangle$ components separately by shutting beams 3 and 4 or 1 and 2 (cf. Fig. \ref{fig:fig1concept}(c), (d)), respectively. The results are shown in Figure \ref{fig:fig2prop}. In Fig. \ref{fig:fig2prop}(a), we show the simulated transverse intensity profile of $|\Psi\rangle$ when a horizontally aligned polariser is positioned in front of the CCD, reflecting the polarisation distribution illustrated in Fig.~\ref{fig:fig0conceptnew}. The normalised intensity profiles for the $|R\rangle$ (beam 1+2) and $|L\rangle$ (beam 3+4) polarisation components are shown in Fig. \ref{fig:fig2prop}(b) and (c), respectively, for different positions $k_zz+\varphi\in[0,\,\pi]$ (arrow at the bottom). For both, $|R\rangle$ and $|L\rangle$ parts, we observe a sinusoidal variation of intensity depending on $k_zz+\varphi$, representing a longitudinal interference pattern of included beams. Furthermore, the variation in intensity for $|R\rangle$ and $|L\rangle$ is out of phase, i.e. $|R\rangle$ components carry maximum intensity while $|L\rangle$ parts are at minimum and vice versa. This behavior is attributed to the phase shift $\alpha_{\text{VB}_{1,2}}$ used to create orthogonally polarised vector beams (cf. Eq. (\ref{eq:PsiClassical})). Moreover, these counter-fluctuating intensities evinces the variation between pure vector and pure scalar beam for $|\Psi\rangle$: If $|R\rangle$ ($|L\rangle$) polarised components are at maximum, while $|L\rangle$ ($|R\rangle$) parts disappear, $|\Psi\rangle$ is represented solely by $|R\rangle$ ($|L\rangle$) components, thus, the light field is purely scalar with $E(|\Psi\rangle, z)=0$, $k_zz+\varphi=\lbrace 0,\,\pi/2,\,\pi\rbrace$. In contrast, if $|R\rangle$ and $|L\rangle$ parts are of equal intensity, $|\Psi\rangle$ is a pure vector mode with $E(|\Psi\rangle, z)=1$, $k_zz+\varphi=\lbrace\pi/4,\,3\pi/4\rbrace$. Between these extreme cases, a smooth transition is found (cf. (a)).\\
\begin{figure}[h]
\centering
\includegraphics[width=.5\textwidth]{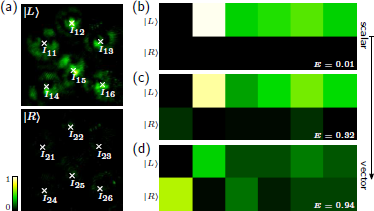} 
\caption{(a) Typical intensity images acquired with a CCD camera to determine the degree of entanglement $E$ showing the case of a scalar beam. The corresponding intensities $I_{uv}$ with $u,\,v\in \lbrace 1,\,2,\,3\rbrace$, arranged according to Table \ref{tab:Tomography} for the cases of a (b) scalar, (c) semi-vector and (d) vector beams, with corresponding values $E=0.05$, $0.54$, and $ 0.96$, respectively.
\label{fig:figTomographyGreen}} 
\end{figure}
\vspace{0.5cm}

\begin{figure*}[thbp]
\centering
\includegraphics[width=\textwidth]{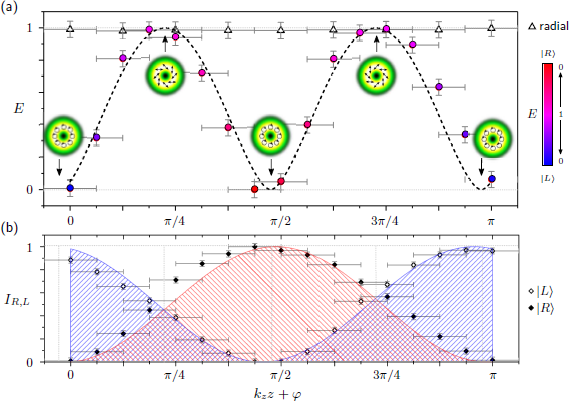} 
\caption{Propagation dynamics of entanglement: (a) Entanglement analysis of pure radial vector mode (black triangles) and created light field $|\Psi\rangle$. Measured $E$ as a function of $k_zz+\varphi$ ($\varphi = -\pi/4$) of the latter is marked by black circles filled according to ratio between $|L\rangle$ and $|R\rangle$ parts (see scale bar). Exemplary modes are shown as green insets. Black dashed curve represents theoretical fit according to Eq. (\ref{eq:nonunitaryDegEnt}). (b) Respective intensity $I_{R,L}$ of $|R\rangle$ (red fit, black filled diamonds) and $|L\rangle$ (blue fit, black hollow diamonds) components of $|\Psi\rangle$ oscillating out of phase. \label{fig:fig3graph}} 
\end{figure*}

\textbf{Entanglement oscillation.} To quantitatively verify the longitudinal entanglement oscillation of $|\Psi\rangle$, we performed an entanglement entropy analysis while digitally propagated the field. By this, we determined the degree of entanglement $E(|\Psi\rangle, z)\in~[0,\,1]$, as a function of $k_zz+\varphi$. The respective experimental method is visualized in Fig. \ref{fig:fig1concept}(c) (blue box) and (e).\\
Figure \ref{fig:figTomographyGreen}(a) shows typical intensity images obtained in experiments per $z$-distance and from which $E$ is computed. The illustrated case corresponds to the scalar field shown in Fig. \ref{fig:figTomographyGreen}(b).  Figures  \ref{fig:figTomographyGreen}(b), (c) and (d) shows the intensity values, normalised and arranged in the form of Table I. Here, we show three cases: scalar, semi-vector and vector beam with corresponding values $E=0.05$, $0.54$ and $0.96$, respectively. The complete set of experimental $E$ values obtained as a function of the propagation distance $z$ are presented in Fig. \ref{fig:fig3graph}. Here, the degree of entanglement (a) and normalised intensity of right-/ left-handed circularly polarized light $I_{R,L}$ (b) are illustrated as a function of $k_zz+\varphi$. Errors of $k_zz+\varphi$ are given by SLM flickering ($\pm \pi/16$), whereby error bars for $E$ ($\pm 0.05$) or $I_{R,L}$ ($\pm 0.03$) are given by inaccuracies within the experimental method/ system.\\
For comparison, we experimentally performed entanglement analysis of a pure radial vector mode (beam 1+3). As theoretically expected (cf. Materials and Methods, Theory), this beam reveals an entanglement entropy of approximately $1$ for all propagation distances, as depicted by black triangles in (a). In contrast, the entanglement dynamics of the engineered beam $|\Psi\rangle$ given by Eq. (\ref{eq:PsiQuantum}), confirms our theoretical predictions, oscillating between pure scalar and pure vector as shown in Fig. \ref{fig:fig3graph}(a). Data is represented by black circles filled according to the ratio between included $|L\rangle$ (blue) and $|R\rangle$ (red) polarised parts (see scale bar). Green insets indicate the modes of light at specific positions. The experimental results reflect the theoretical description in Eq. (\ref{eq:nonunitaryDegEnt}) with $k_zz$ replaced by $k_zz+\varphi'$ perfectly, as illustrated by the according fit in (a) (black dashed curve). The fitting parameter $\varphi'$ has a value of $-0.71$, thus, almost matches chosen setting of $\varphi=-\pi/4$. \\
Figure \ref{fig:fig3graph}(b) shows simultaneously determined counter-fluctuating intensity curves for $|L\rangle$ (blue fit, black hollow diamonds) and $|R\rangle$ (red fit, black filled diamonds). Obviously, these curves mirror the propagation dynamics of entanglement and the involved variation in ratio between $|L\rangle$ and $|R\rangle$ demonstrated in (a). A slight shift with respect to positions of extrema of $|L\rangle$ and $|R\rangle$ can be observed which reflects the deviation between $\varphi$ and $\varphi'$. Our results prove that by adjusting $\varphi$ it is possible to transport a desired degree of entanglement in $|\Psi\rangle$ to a predefined $z$-position. \\

Our results highlight the fact that it is possible to engineer vectorial light fields whose degree of non-separability oscillates in free-space, from fully vector to fully scalar, as function of the propagation distance. While we have restricted ourselves to first-order vector vortex beams for the demonstration, the concept we outline here is more general and can be applied to higher-order vortex modes as well as, in principle, any vector state with judicious choice of degree of freedom.  
The surprising result is that our entanglement dynamics take place in free-space under unitary conditions.  We emphasis that while we have performed our experiments with coherent states for convenience, the same results are obtained for local entanglement of internal degrees of freedom of a single photon.  Neither theory nor experiments differentiate between these two cases, thus, addressing hot questions as to the notion of local and classical entanglement and its propagation dynamics.

An important aspect of this work is the practical approach to the generation and propagation of the fields. It is possible to engineer the desired effect using a Sagnac interferometer in which an input radially polarised vector beam is split into two beams traveling along each arm: one of the beams is switched to azimuthal polarisation, with a half-wave plate, and interfered with the radially polarised beam. In the third arm, both beams propagate in opposite directions bearing orthogonal states of polarisation, generating in this way a standing wave whose degree of entanglement varies along the propagation axis. This generating approach does not allow one to experimentally measure the entanglement nor to deliver a desired state to some location. We offer a more powerful approach that utilises digital generation and propagation enabled by a SLM. This approach allowed us to manipulate each vector beam independently and among other things to perform digital propagation on each. Hence both vector beams propagate in a collinear fashion in a manner that simulates propagation in opposite directions. This approach of generation and propagation enabled us to realise a vector field with a non-constant degree of entanglement that can be changed in time and space by simply changing the displayed hologram. Importantly, this approach allowed us not only to monitor the degree of non-separability but also to deliver specific states to arbitrary positions along the propagation direction. We believe this approach and result will broaden the applications of vector beams, for example, by adjusting in real time the vector state of light at a target plane for laser materials processing, say from circularly polarised to radially with only a hologram change, within a small volume for micro-fluidic control, and by adjusting the mode at the focal region in STED systems.  All would benefit from a virtual propagation of co-propagating modes of the type described here. 

\section*{Conclusion}
We have demonstrated that by exploiting complex modes of light it is possible to have an oscillating degree of local entanglement during propagation, even though the medium is considered unitary, i.e., a medium where the entanglement should not change.  In essence we have created the first tractor beam for local entanglement, able to deliver a known degree of entanglement to some target plane.  We have shown this with entangled internal degrees of freedom of polarisation and spatial mode, and while our experiment was classical the results hold equally well for local entanglement of internal degrees of freedom of a single photon.  Our approach highlights intriguing questions about the notion of entanglement dynamics and offers a new tool for the delivery of controlled vector states to targets, which we believe will be of practical value to a range of communities, from materials processing to imaging.

\section*{Materials and correspondence}
Correspondence and requests for materials should be addressed to C.R.G. 


\section*{Acknowledgments}
E.O. acknowledges financial support from the German Research Foundation DFG (EXC 1003 -– CiM, TRR61) C.R.G from the Claude Leon foundation and CONACyT and B.N. from the National Research Foundation of South Africa.  The authors would like to thank Thomas Konrad for useful advice.



\section*{Authors' contributions}
Experiments were performed by E.O. and C.R. with theoretical input from B.N.  All authors contributed to the data analysis, interpretation of the results and writing of the manuscript.  A.F. conceived the idea and supervised the project.



\section*{Competing financial interests}
The authors declare no financial competing interests.\\

\newpage
\bibliographystyle{apsrev4-1}
\bibliography{refs}

\newpage \newpage\newpage
\section*{Supplementary information \label{sec:Supplement}}
\subsection*{Superposition of counter-propagating orthogonal vector fields -- Theory\label{Ssubsec:SuperposOrthoVectorFields}}

The realisation of the light field $|\Psi(x,y,z)\rangle$ with a spatially varying degree of entanglement $E(|\Psi\rangle, z)$ was obtained by combining two orthogonal vector beams VB$_1$ and VB$_2$ propagating in opposite $z$-directions. These vector modes represented by $|\Psi_{\text{VB}_1}^{+}\rangle$ and $|\Psi_{\text{VB}_2}^{-}\rangle$ are generated by setting $\alpha_{\text{VB}_1}=0$ and $\alpha_{\text{VB}_2}=\pi/2$, respectively, with $l=l_1=-l_2$ and $p=p_1=p_2$ (cf. Eq. (\ref{eq:VectorBeams})). The resulting light field of the superposition can be written as
\begin{align}
|\Psi\rangle &= \frac{1}{\sqrt{2}}\left(|\Psi_{\text{VB}_1}^{+}\rangle+|\Psi_{\text{VB}_2}^{-}\rangle\right) \nonumber\\
&=\frac{1}{2}\left(|LG_{p}^{-l}\rangle\,|R\rangle+|LG_{p}^{l}\rangle\, |L\rangle\right)\cdot \text{e}^{\text{i}k_zz} \nonumber\\
&\,\,\,+\frac{1}{2}\left(\text{e}^{\text{i}\frac{\pi}{2}}\,|LG_{p}^{-l}\rangle\,|R\rangle+\text{e}^{-\text{i}\frac{\pi}{2}}\,|LG_{p}^{l}\rangle \,|L\rangle\right)\cdot \text{e}^{-\text{i}k_zz}.
\label{eq:PsiClassical}
\end{align}
Regrouping terms with same polarisation leads to
\begin{align}
|\Psi\rangle = &\frac{1}{2}\left(\text{e}^{\text{i}k_zz}+\text{i}\,\text{e}^{-\text{i}k_zz}\right)\,|LG_{p}^{-l}\rangle|R\rangle \nonumber\\
&+ \frac{1}{2}\left(\text{e}^{\text{i}k_zz}-\text{i}\,\text{e}^{-\text{i}k_zz}\right)\,|LG_{p}^{l}\rangle|L\rangle
\label{Seq:PsiQuantum}
\end{align}

\subsection*{Non-separability in orthogonal superpositions of vector fields -- Theory\label{Ssubsec:NonseparabilitySuperposVectorFields}}
The degree of non-separability of a vector field given by
\begin{equation}
|\Psi\rangle = \sqrt{a}\cdot |u_R\rangle|R\rangle + \sqrt{1-a}\cdot |u_L\rangle|L\rangle,
\label{Seq:VBQuantumNotation}
\end{equation}
can be computed as\,\cite{McLaren2015}
\begin{equation}
E(|\Psi\rangle) = -\left[a\cdot \log_2(a)+(1-a)\cdot \log_2(1-a)\right].
\label{Seq:vectorness}
\end{equation}
Comparing Eqs. (\ref{Seq:PsiQuantum}) and (\ref{Seq:VBQuantumNotation}), considering that $|u_{R,L}\rangle = |LG_{p}^{l}\rangle \cdot \text{e}^{\text{i}\zeta_{R,L}}$, one can see that
\begin{align}
\sqrt{a}\cdot \text{e}^{\text{i}\zeta_{R}}=&\frac{1}{2}\left(\text{e}^{\text{i}k_zz}+\text{i}\,\text{e}^{-\text{i}k_zz}\right)\nonumber\\
=& \frac{1}{2} \left[\cos(k_zz)+\sin(k_zz)\right]+\nonumber\\
&\text{i}\frac{1}{2}\left[\cos(k_zz)+\sin(k_zz)\right].
\end{align}
Hence, $\zeta_{R} = \pi/4$ ($=-\zeta_{L}$) and
\begin{align}
\sqrt{a} &= \biggl|\frac{1}{2}\left(\text{e}^{\text{i}k_zz}+\text{i}\,\text{e}^{-\text{i}k_zz}\right)\biggl|\nonumber\\
a&=\frac{1}{4}\left(\text{e}^{\text{i}k_zz}+\text{i}\,\text{e}^{-\text{i}k_zz}\right)\cdot\left(\text{e}^{\text{i}k_zz}+\text{i}\,\text{e}^{-\text{i}k_zz}\right)^*\nonumber\\
&=\frac{1}{4}\left(\text{e}^{\text{i}k_zz}+\text{i}\,\text{e}^{-\text{i}k_zz}\right)\cdot\left(\text{e}^{-\text{i}k_zz}-\text{i}\,\text{e}^{\text{i}k_zz}\right)\nonumber\\
&=\frac{1}{4}\left[2- \text{i}(\text{e}^{\text{i}2k_zz}-\text{e}^{-\text{i}2k_zz})\right]\nonumber\\
&=\frac{1}{4}\{2-\text{i}(2\text{i}\sin(2k_zz))\}\nonumber\\
&=\frac{1}{2}[1+\sin(2k_zz)].
\end{align}
Substitution of $a$ into Eq. (\ref{Seq:vectorness}) yields
\begin{align}
E(|\Psi\rangle, z) &= -\frac{1}{2}[1+\sin(2k_zz)] \log_2\left\{\frac{1}{2}[1+\sin(2k_zz)]\right\}-\nonumber\\
&\frac{1}{2}[1-\sin(2k_zz)])\log_2\left\{\frac{1}{2}[1-\sin(2k_zz)]\right\}.
\end{align}
The above equation can be finally written as
\begin{align}
E(|\Psi\rangle,z) = &1-\frac{1}{2}\left[1+\sin(2k_zz)\right]\cdot \log_2\left[1+\sin(2k_zz)\right]- \nonumber\\
&\frac{1}{2}\left[1-\sin(2k_zz)\right]\cdot \log_2\left[1-\sin(2k_zz)\right].
\end{align}

\subsection*{Counter-propagating higher-order vector modes\label{Ssubsec:HigherorderModes}}
\begin{figure}[hbp]
\centering
\includegraphics[width=.4\textwidth]{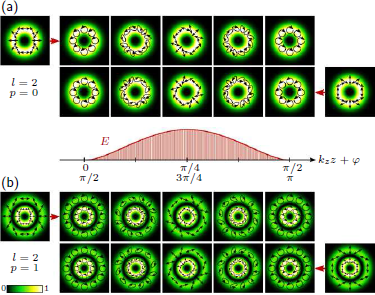} 
\caption{Space-variant degree of entanglement by counter-propagating vector modes of higher order with (a) $l=2$, $p=0$ and (b) $l=2$, $p=1$. Initial modes are shown at the right and left edge, whereby central images present different transverse planes (intensity + polarisation) within $|\Psi\rangle$. Respective degree of entanglement $E$ depending on $k_zz+\varphi$ is depicted by red curve. \label{fig:fig5higherorder}} 
\end{figure}
The realisation of spatially varying degree of entanglement of the form $E(|\Psi\rangle,z)$ is not only facilitated by first-order vector modes with $l=1$ and $p=0$ but also by the application of higher-order modes\,\cite{Otte2016} with $|l|>1$ and $p>0$. In this case, vector modes $|\Psi_{\text{VB}_1}^{+}\rangle$ and $|\Psi_{\text{VB}_2}^{-}\rangle$ still represent counter-propagating modes, whereby $l=l_1=-l_2$ and $p=p_1=p_2$ (cf. Eq. (\ref{eq:VectorBeams})) for both vector modes, as indicated within the theoretical description above. If these requirements are fulfilled, different light fields $|\Psi\rangle$ according to Eq. (\ref{eq:PsiQuantum}) can be created.  Figure~\ref{fig:fig5higherorder} sketches two examples of these fields with (a) $l=2$, $p=0$ and (b) $l=2$, $p=1$. Here, different transverse planes of $|\Psi\rangle$ (normalised intensity + polarisation) are illustrated for chosen propagation distances $z$. The respective degree of entanglement $E$ is visualized by the red curve between (a) and (b), whereby the arrow shows corresponding values of $k_zz+\varphi$ with $[0,\,\pi/2]$ and $[\pi/2,\,\pi]$  belonging to first and second line of (a) and (b). Initial vector modes propagating in $+z$- and $-z$-direction are indicated at the left and right edge, respectively. Note that, function $E(|\Psi\rangle, z)$ is independent of chosen mode numbers $l$ and $p$, even if other characteristics as intensity and polarisation of respective light field $|\Psi\rangle$ change according to $l$ and $p$.    

\end{document}